\documentclass[]{spie}  

\usepackage{amsmath,amsfonts,amssymb}
\usepackage{graphicx}
\usepackage[colorlinks=true, allcolors=blue]{hyperref}

\title{AAO Starbugs: software control and associated algorithms}

\author[a]{Nuria P. F. Lorente}
\author[a]{Minh V. Vuong}
\author[a]{Keith Shortridge}
\author[a]{Tony J. Farrell}
\author[a]{Scott~Smedley}
\author[b]{Sungwook E. Hong}
\author[a]{Carlos Bacigalupo}
\author[a]{Michael Goodwin}
\author[a]{Kyler Kuehn}
\author[a,c]{Christophe~Satorre}
\affil[a]{Australian Astronomical Observatory, PO Box 915, North Ryde, NSW 1670, Australia}
\affil[b]{School of Physics, Korea Institute for Advanced Study, Seoul 02455, Korea}
\affil[c]{Laboratoire de Syst\`{e}mes Robotiques (LSRO), Ecole Polytechnique F\'{e}d\'{e}rale de Lausanne, Switzerland}

\authorinfo{N.P.F.L: e-mail: nuria.lorente@aao.gov.au; telephone: 61 2 9372 4897; \url{www.aao.gov.au}}

\begin{document} 
\maketitle

\begin{abstract}
The Australian Astronomical Observatory's TAIPAN instrument deploys 150 Starbug robots to position optical fibres to accuracies of 0.3~arcsec, on a 32~cm glass field plate on the focal plane of the 1.2~m UK-Schmidt telescope. This paper describes the software system developed to control and monitor the Starbugs, with particular emphasis on the automated path-finding algorithms, and the metrology software which keeps track of the position and motion of individual Starbugs as they independently move in a crowded field.
The software employs a tiered approach to find a collision-free path for every Starbug, from its current position to its target location. This consists of three path-finding stages of increasing complexity and computational cost. For each Starbug a path is attempted using a simple method. If unsuccessful, subsequently more complex (and expensive) methods are tried until a valid path is found or the target is flagged as unreachable.
\end{abstract}


\section{INTRODUCTION}
\label{sec:intro}  

Multi-Object fibre Spectroscopy (MOS) is a well-established technique for 
carrying out spectroscopy on a large number of targets in the field of view. 
It attempts to solve the problem of focal plane inefficiency, aiming to use the full field of the telescope when the object of interest occupies a small fraction of the available focal surface.
The advent of optical fibres brought us several ways of improving our use of telescope collecting area: now we can selectively pick-off light from a portion of our aperture, and multiplex this to the number of fibres available. 
The challenge, then, is to build a positioner to accurately, quickly and efficiently place the optical fibres in the desired location on the optical plane,  corresponding to the position of the object of interest in the sky. 

There have been a number of techniques successfully employed to do this in the last couple of decades:
The 2dF positioner\cite{2002lct+}, consists of a robot arm which, in sequence, picks up and positions each of 400 fibres onto the Anglo-Australian Telescope's 560~mm field plate. The now decommissioned 6dF\cite{2001ASPC..232..421W} at the UKST worked in a similar way, positioning 150~fibres onto a 320~mm plate giving a $6^\circ$ field of view (FoV).
A different type of positioner, {Echidna\cite{2016gd}}, makes use of rigid spines, each housing a fibre and pivoting on a bearing. 
A multi object field can also be achieved using plug-plates\cite{2015MNRAS.447.2857B} --- this may sound primitive, but it has the advantage of allowing versatility in payload size and packaging, is useful for rapid prototyping, as well as being a lower cost option. 
The disadvantages include long reconfiguration times, and zero target flexibility once the plug-plates have been fabricated.

One problem with traditional pick-and-place positioners is the time they take to configure the field. As an example, 2dF takes around 40~minutes to configure its $2^\circ$ field, because of its single arm and the serial nature of its operation. Additionally, field configuration time increases linearly with number of fibres.
This makes it impractical to use this type of positioning instruments for studies which require short observations (e.g. bright stellar targets) where the field reconfiguration time may be an order of magnitude longer than the exposure time.
Other limitations include an inability to change the payload (e.g. to exchange the single fibre payload for a fibre bundle), or to manage a non-planar focal plane.
To overcome these shortcomings the AAO has been developing the Starbugs technology, which exchanges the single-armed robot with one independently positionable robot per science fibre. 

\section{STARBUGS}

\begin{figure} [ht]
\begin{center}
\begin{tabular}{cc} 
\includegraphics[width=0.44\linewidth]{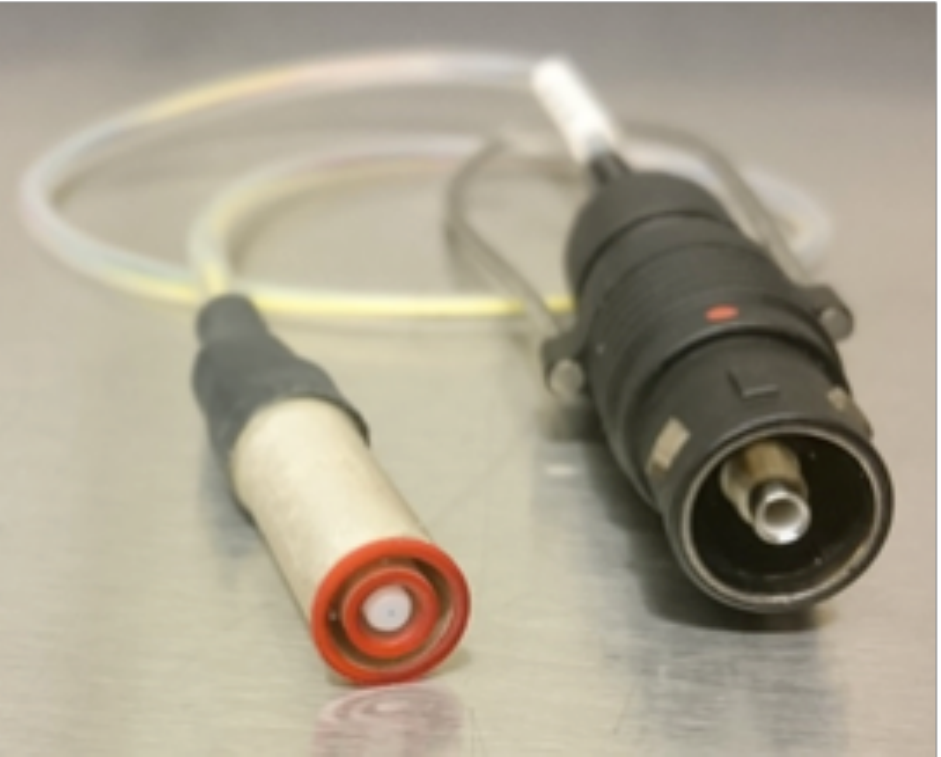} &
\includegraphics[width=0.485\linewidth]{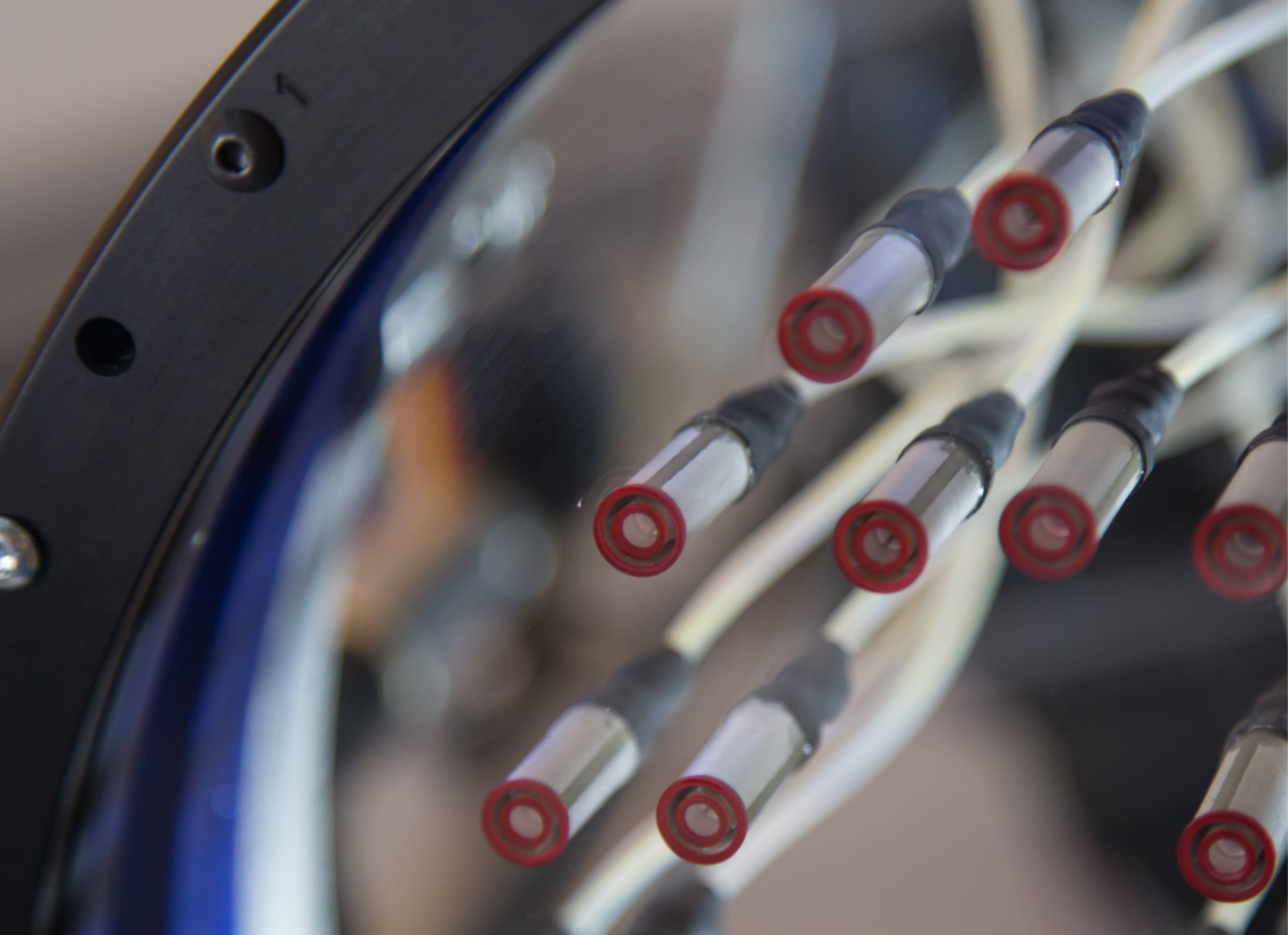}
\end{tabular}
\end{center}
\caption[starbug] 
{ \label{fig:starbug} 
Left: A starbug (left) with its umbilical and connector (right). The end of each piezo-ceramic tube is encased with a red glass-like, plastic slipper providing rigidity to the Starbug tubes and increased friction against the field plate to facilitate motion, and minimise abrasion damage to the field plate. Right: A number of Starbugs are shown deployed on the glass field plate of the TAIPAN instrument.}
\end{figure}

Starbugs\cite{2016slb+} are 9~mm diameter, individually positionable robots composed of two concentric piezo-ceramic tubes (see Figure~\ref{fig:starbug}, left). These vibrate when a voltage is applied across electrodes located on opposing sides of the inner and outer tubes, which results in a walking motion when the Starbugs are placed on a glass field plate located at the telescope's optical surface (Figure~\ref{fig:starbug}, right).
Each Starbug  has an umbilical which carries power, vacuum, 3 metrology fibres, 
and a science payload, which for the first instrument using this technology is a single science-grade fibre (the use of other payloads such as hexabundles and lenslets, is under development).

Starbugs provide a distinct advantage over single robot-arm positioners: each robot can be individually controlled and positioned, so all Starbugs can be reconfigured at once. In the case of the TAIPAN instrument, this translates to a reconfiguration time of less than 5~minutes, making feasible studies which were previously considered inefficient.
Because the Starbugs are moved in parallel, the  linear relationship between reconfiguration time and number of science fibres is therefore greatly flattened.
Finally, because the Starbugs are self-propelled and small, they can be used on a curved focal plane.

\section{The TAIPAN Instrument}
TAIPAN is a multi-object fibre-fed Starbug positioner\cite{2016slb+} and dedicated dual-band spectrograph\cite{2016slz+} for the newly-refurbished 1.2~m UK-Schmidt Telescope at Siding Spring Observatory. 
In the first instance TAIPAN will deploy 150 Starbugs on its 32~cm glass field plate,
with plans to upgrade this to 300 Starbugs in the first year of operations. The instrument will have FoV = $6^\circ$, with FoV = 3.3~arcsec for the individual Starbugs.

The TAIPAN instrument will be used by two science surveys over the next 5 years: 
\begin{itemize}
\item The {\bf Taipan} survey aims to measure the spectra for $0.5-1 \times 10^6$ galaxies in the southern hemisphere to $i<18$ with $R=1960$ (blue) to $R=2740$ (red), and  $5\leq\,$SNR\,$\leq10$.
The survey will measure the present-day expansion rate of the Universe, $H_0$, and the growth rate of structure to 1\% and 5\% precision respectively.
It will also carry out a peculiar velocity survey, which will make the most extensive map to date of the mass distribution and motions in the local Universe.
Finally, the survey will work to understand the role of mass and environment in the evolution of galaxies.

\item The {\bf FunnelWeb} stellar survey will characterise the brightest $(I<12)$ $3\times10^6$ stars in our Milky Way Galaxy with SNR~$\sim100$ and $R\sim 2000$, to produce a spectral library of stellar properties, including effective temperature ($T_{\mathrm{eff}}$), surface gravity ($log(g)$), metallicity ([Fe/H]), and alpha element to iron abundances ($\mathrm{\alpha/Fe}$). Additionally, the survey will inform future studies of exoplanets that may orbit those stars.

\end{itemize}

\section{The TAIPAN Software System}
\begin{figure} [ht]
\begin{center}
\includegraphics[width=\linewidth]{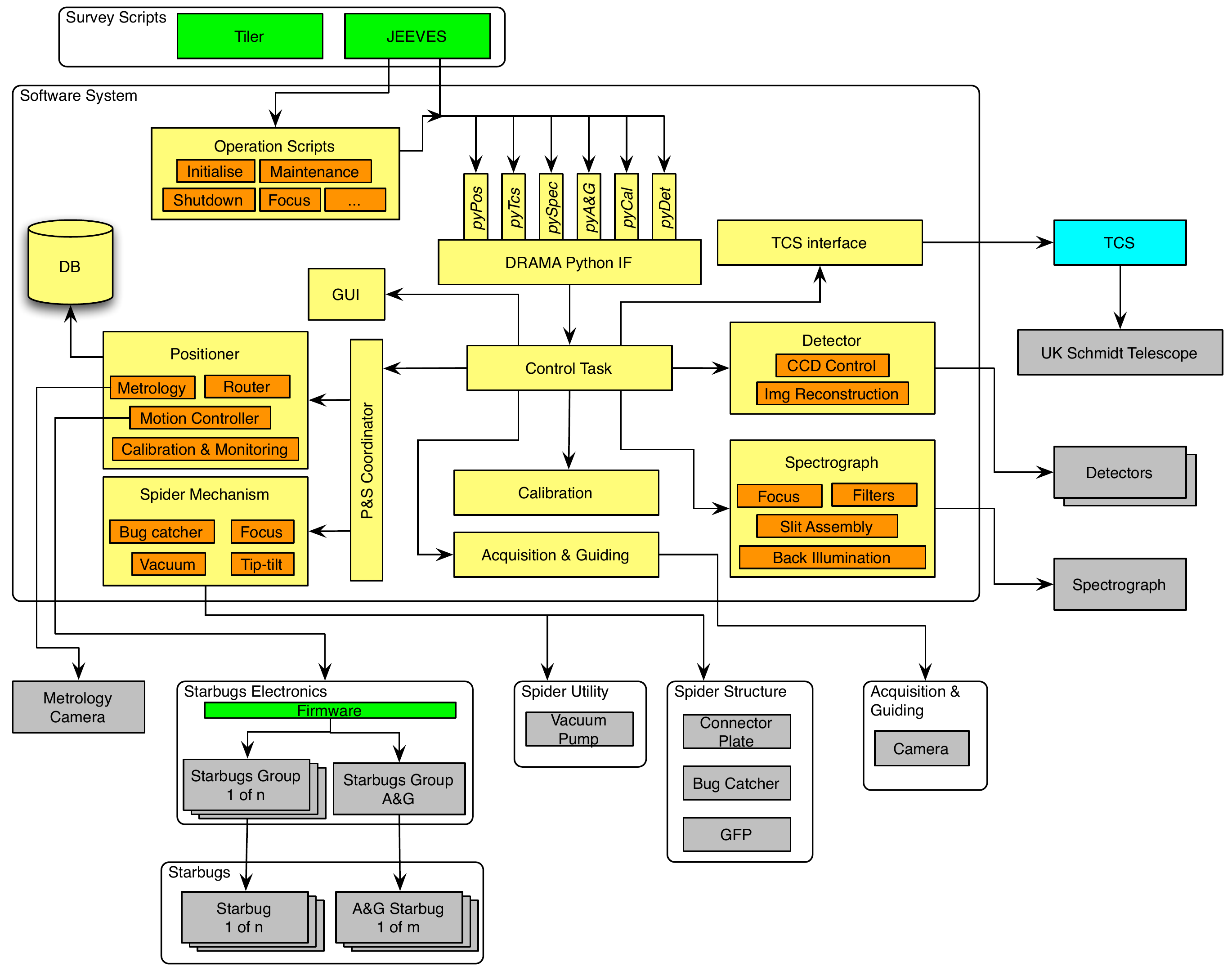}
\end{center}
\caption[systemDiagram] 
{ \label{fig:systemDiagram} 
System diagram showing the modules comprising the TAIPAN software system (yellow and orange), and the relationships between them, in the context of the instrument's electronics and hardware (gray) and other software modules (green and blue).}
\end{figure}

The TAIPAN software system is responsible for the control and monitoring of the instrument's positioner, spider mechanism, spectrograph, and detectors, and its calibration and acquisition \& guiding modules. It also interfaces to the telescope via a third-party Telescope Control System (TCS), and to the survey team's JEEVES sequencer --- a set of python scripts which will autonomously schedule and run each night's observations based on a set of pre-computed survey fields.
Figure~\ref{fig:systemDiagram} shows a schematic of the software system in the context of the telescope, instrument hardware and electronics, and the survey team's tiling, scheduling and data pipeline software.

\subsection{Middleware: DRAMA~2}
The TAIPAN software system uses the new DRAMA~2\cite{2015fs} middleware API for coordinating actions and inter-process communication.
Since it was developed in 1992 DRAMA\cite{1995fbs} has been the AAO's primary middleware tool for constructing complex instrumentation systems. It uses a tasking model where tasks are potentially running on different hosts in a heterogeneous environment. A DRAMA task responds to command messages and executes defined actions. Tasks can co-operatively multi-task: that is, several actions may run simultaneously, returning control to the main DRAMA loop between events so as to allow other actions to run, or for the action itself to receive a new message.

Because of its cooperative multi-tasking, it has not been necessary for DRAMA to support multi-threading itself (although DRAMA tasks which make use of threads have been developed over the years). 
However, as threads have become more thoroughly supported by C++ compilers and standard libraries (especially since the advent of C++11), 
the time was right to implement DRAMA~2, which simplifies the writing and maintenance of complex DRAMA tasks.

We have chosen to use DRAMA 2 for the TAIPAN instrument as a modern alternative to DRAMA. DRAMA~2 fully supports threading and allows reliable tasks to be written quickly. It facilitates the development of sequenced code without losing the efficient, non-blocking, DRAMA messaging facilities. From the development point of view, DRAMA 2 also hides much of the complexity of creating threaded distributed applications, thereby lowering the associated risk.

As well as being a science-driven instrument in its own right, TAIPAN serves as a technology prototype for the MANIFEST robotic fibre positioning system designed for the Giant Magellan Telescope\cite{2016lbb+}.
As such, we have thought it prudent to design and develop the software system with a mind to its reuse on MANIFEST. 
Several of the modules have been designed so as to keep the DRAMA~2 and business parts of the code in separate sub-modules, 
to facilitate the replacement of DRAMA~2 with whichever middleware is selected for use on GMT.

\begin{figure} [b]
\begin{center}
\begin{tabular}{cc} 
\includegraphics[height=70 mm]{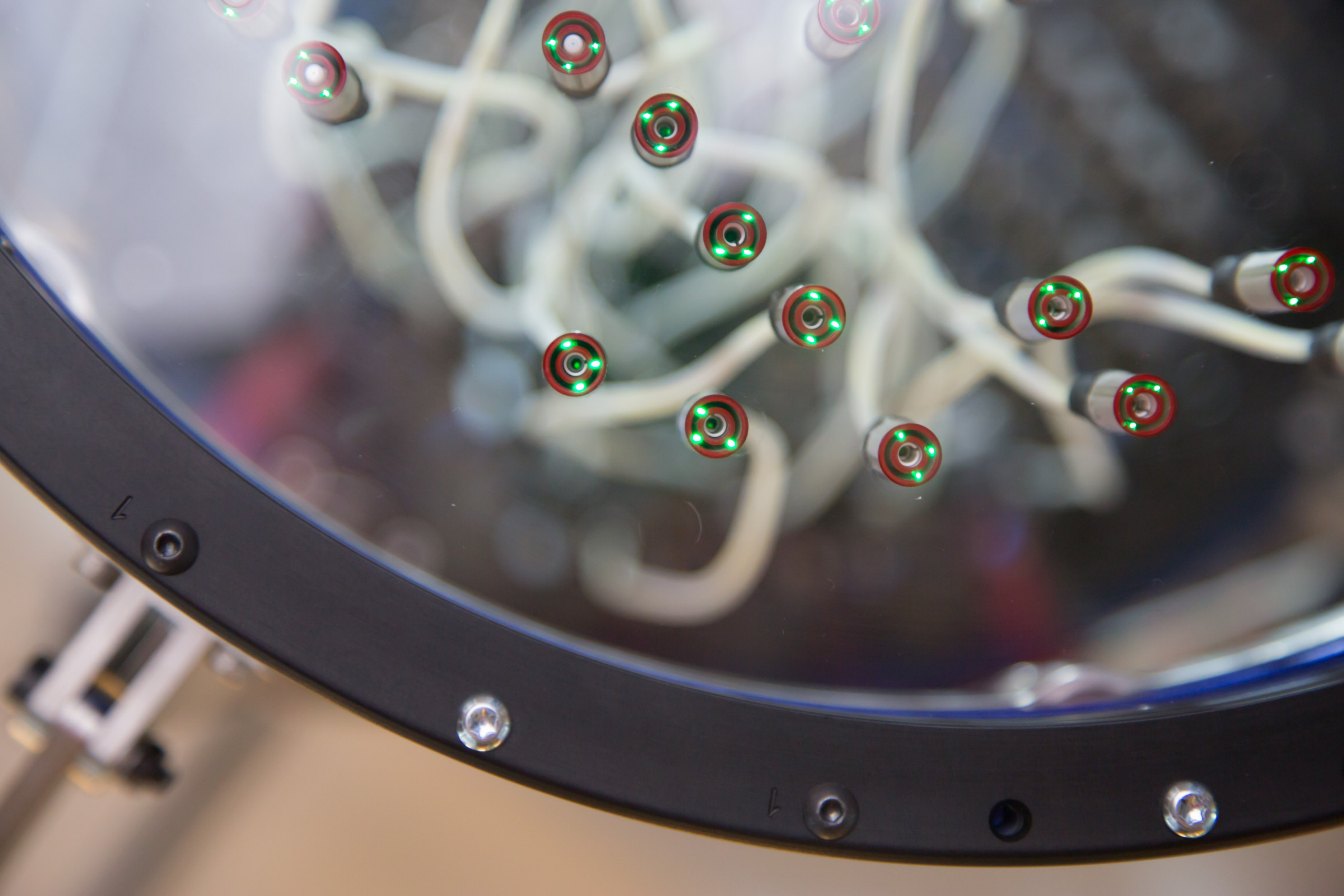} &
\includegraphics[height=70 mm]{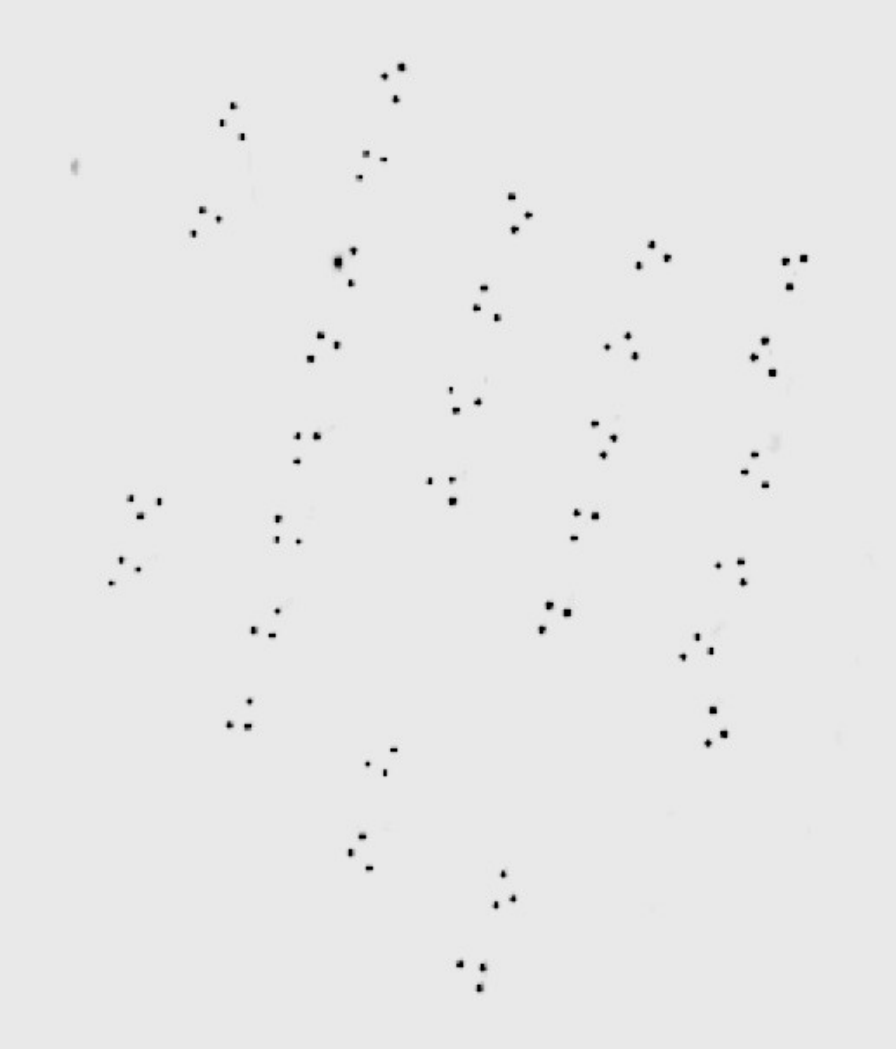}
\end{tabular}
\end{center}
\caption[starbugMetFibres] 
{ \label{fig:starbugMetFibres} 
Left: Starbugs with illuminated metrology fibres, viewed from the perspective of the metrology camera, located underneath the glass field plate. Right: A portion of the metrology image, with enhanced contrast and inverted greyscale to improve clarity.}
\end{figure} 

\subsection{Metrology}
The Metrology software subsystem controls the 29~M-pixel camera which images the triplet of back-illuminated metrology fibres located on each Starbug (Figure~\ref{fig:starbugMetFibres}). It processes the resulting $6576 \times 4384$~pixel images to find the centroid of each metrology fibre, and from this calculate the position of the geometric centre of the Starbug, and of the central science fibre.
The Metrology subsystem uses the AAO's generic Fibre Metrology software package, which supports single and multiple metrology fibre system, and is used to carry out metrology across a number of AAO instruments.

The Metrology subsystem forms part of the Starbug control closed loop (together with the Positioner). Running at 4~Hz, this keeps track of the positions of all the Starbugs during the reconfiguration, detects `missing' Starbugs: this can happen if one or more of a Starbug's metrology fibres fail to illuminate, if a Starbug's contact with the plate is at an angle other than $90^\circ$, or if a Starbug falls off the field plate due to a vacuum failure, etc. The Metrology subsystem also helps to diagnose unusable Starbugs --- e.g. one which does not respond to movement command, has faulty metrology fibres, or has otherwise been marked as bad by the system. This subsystem is also crucial in Starbug initialisation and position calibration.

\subsubsection{Metrology Camera Calibration}
In the ideal case the plate cartesian coordinates of the Starbugs can be directly calculated from their pixel positions by simply applying the plate scale magnification factor.
In reality, however, the mapping between the x/y-coordinates and pixel positions is not so linear, for two principal reasons.
First, the focal plate to which the  Starbugs are attached may not be perfectly aligned with respect to the metrology camera ({\em plate misalignment}).
Second, the metrology camera itself distorts the image ({\em camera distortion}).

To characterise the plate misalignment we used five free parameters --- x/y/z-directional rotations and x/y-directional offsets --- by assuming a fixed distance between the focal plate and the metrology camera.
For characterising the camera distortion, on the other hand, we adopted the model in {\em Camera Calibration Toolbox for Matlab}\footnote{http://www.vision.caltech.edu/bouguetj/calib\_doc/} with the following assumptions:
(1) the angle between the x- and y-pixel axes is $90^{\circ}$;
(2) the focal length is assumed to be identical over direction; and
(3) the $r^6$-order radial distortion term is set to zero, because allowing a nonzero value often leads to the wrong estimation of the overall camera distortion parameters.

\begin{figure} [ht]
\begin{center}
\begin{tabular}{cc} 
\includegraphics[height=80 mm]{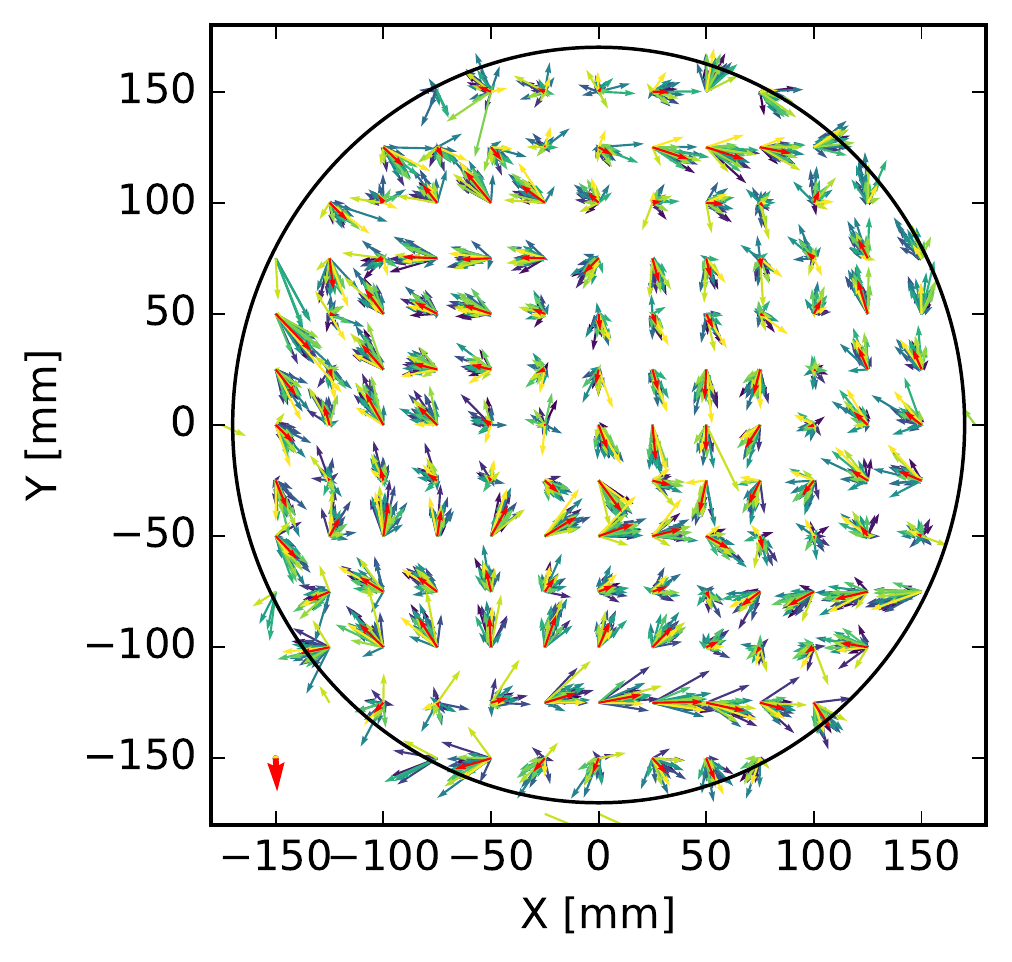} &
\includegraphics[height=82 mm]{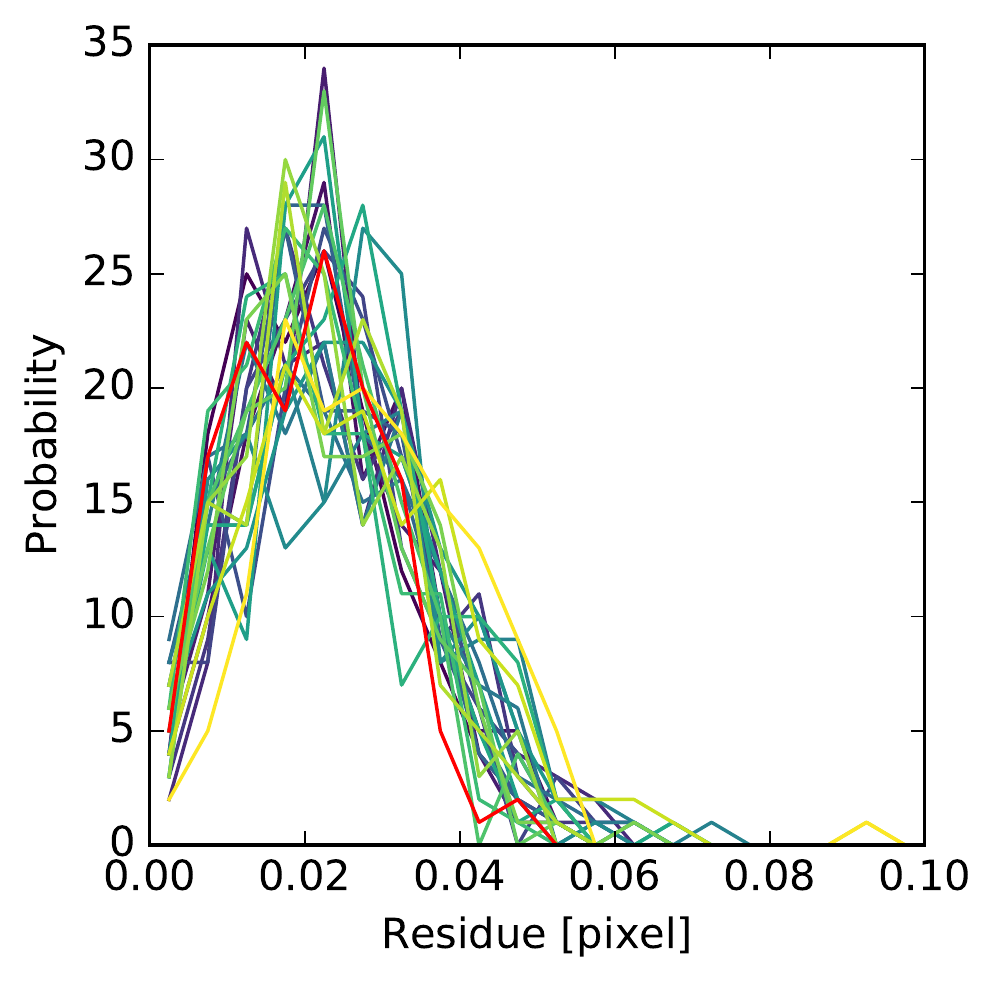}
\end{tabular}
\end{center}
\caption[starbugMet] 
{ \label{fig:starbugMet} 
Results of lab measurements of a plane plate with a 25 mm pitch uniform grid with 25 different plate distortion setups, for the estimation of camera distortion parameters.
Left: Residual shape of position estimation from 25 setups (yellow to green) and their average (red).
The black circle shows the TAIPAN FoV.
The length of the arrows within the FoV is magnified 500 times for visibility.
Right: Distribution of residue size from 25 setups (yellow to green) and their average (red).
RMS $\sim 0.02$~pixels, which is $\sim2$~microns.}
\end{figure} 

Figure~\ref{fig:starbugMet} shows the results of the lab measurements of the metrology camera's  distortion parameters.
The object in these measurements is the donut-dot chrome glass grid manufactured by Max Levy Autograph, Inc., which has donut-shaped dots within a uniform 25 mm-pitch grid.
We carried out 25 different plate distortion tests, and took 10 frames and averaged the positions of the centroid for each.
The best-fit camera distortion parameters were then estimated, by allowing both the camera distortion parameters, and 25 sets of plate distortion parameters to be free parameters.
The RMS of the difference between the actual grid positions and their estimations by the metrology camera is $\sim2$~microns, which is well within the 4.5~micron (0.3~arcsec) TAIPAN requirement.

During observations, the alignment between the metrology camera and the focal plate may change due to telescope movement.
To estimate the plate misalignment, we first assume the camera distortion parameters found in the lab measurements and find the best-fit plate misalignment parameters by using the {\em plate fiducials}: 17 fixed fibres located at the predetermined positions around the focal plate.
Then, if necessary, the camera distortion parameters can be recalibrated by assuming the above plate misalignment parameters, etc.
From the simulation, we found that 16 plate fiducials are sufficient to reproduce the x-/y-coordinates of Starbugs fibres within a 1-micron error, once we assume the correct camera distortion model.

\subsection{Starbug Personalities}
Because Starbugs walk by virtue of the vibration of their piezo-ceramic tubes, their motions are not ideal and not always as expected.
Several factors affect the direction and speed of a Starbug; these include asymmetries in the positions of the electrodes on the inner and outer tubes, which are expected to give consistent deviations from ideal motion over the lifetime of the Starbug. Less obviously, dust and debris on the field plate also affect the motion, as a function of position on the plate and how long ago the plate was last cleaned. Finally, the (random) interaction between a Starbug's umbilical cord and that of its neighbouring Starbugs can have a surprisingly large effect which is difficult to eliminate or quantify as it is governed by the relative positions of the Starbugs on the plate as well as the elevation angle of the telescope.
The closed-loop software control compensates for these factors, ensuring that the Starbugs reach their target positions.

\begin{figure} [hb]
\begin{center}
\begin{tabular}{cc} 
\includegraphics[height=80 mm]{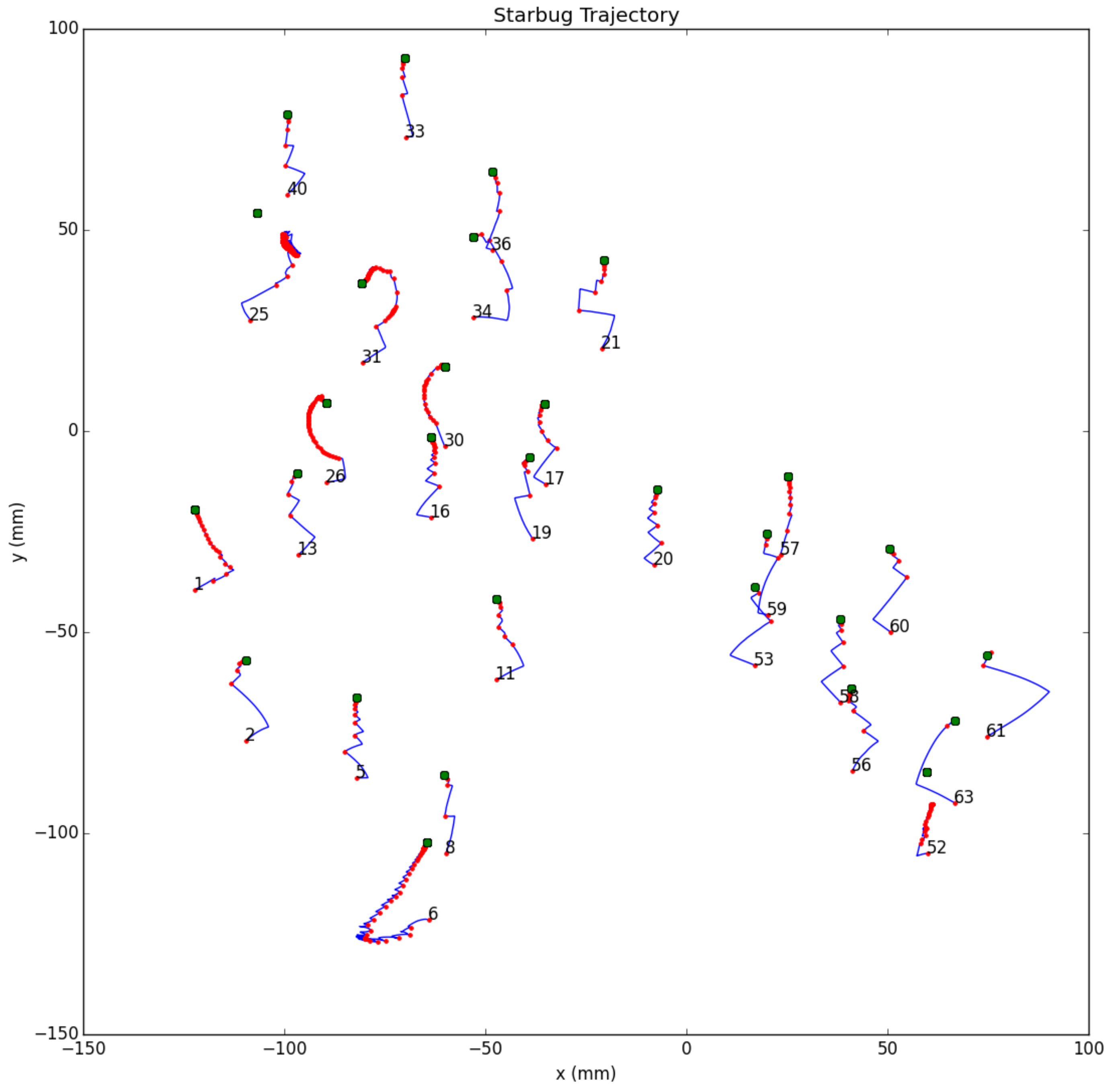} &
\includegraphics[height=80 mm]{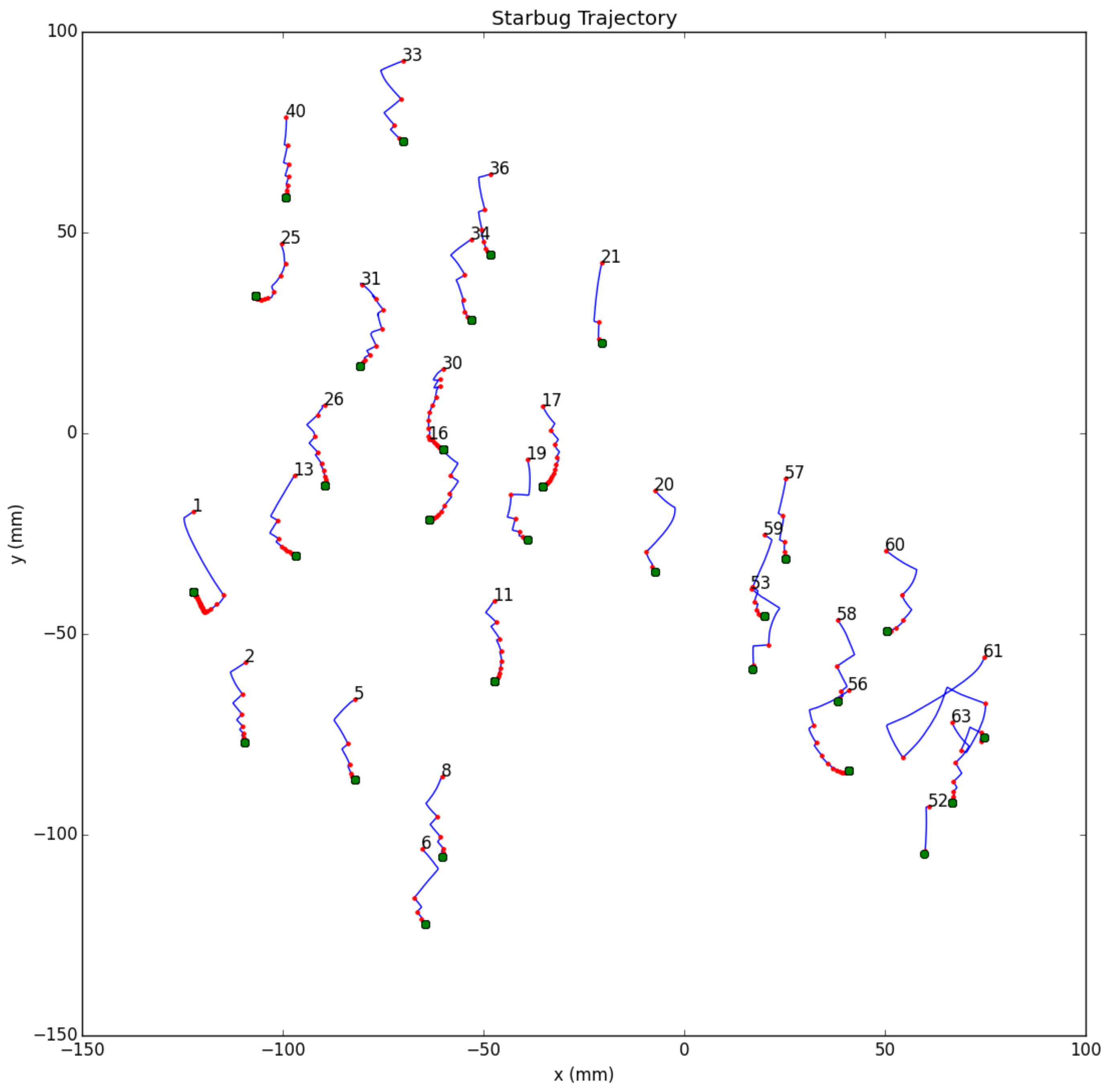}
\end{tabular}
\end{center}
\caption[sbPersonalities] 
{ \label{fig:sbPersonalities} 
The positions of 29 Starbugs during the course of a translational motion test (i.e. no rotation was employed). The Starbugs were under closed-loop control and were not calibrated for step size or angular deviation from ideal motion. Each trajectory begins on a numbered red dot (the number is the Starbug's ID) and the requested destination is marked by a green dot. The actual trajectory is marked by a blue line, with subsequent red dots marking a course correction control point.
The Starbugs began the test facing in random orientations and were then commanded to move to a position 20~mm in the vertical direction (left) and to return to their original positions (right).}
\end{figure}

\begin{figure} [ht]
\begin{center}
\includegraphics[width=0.7\linewidth]{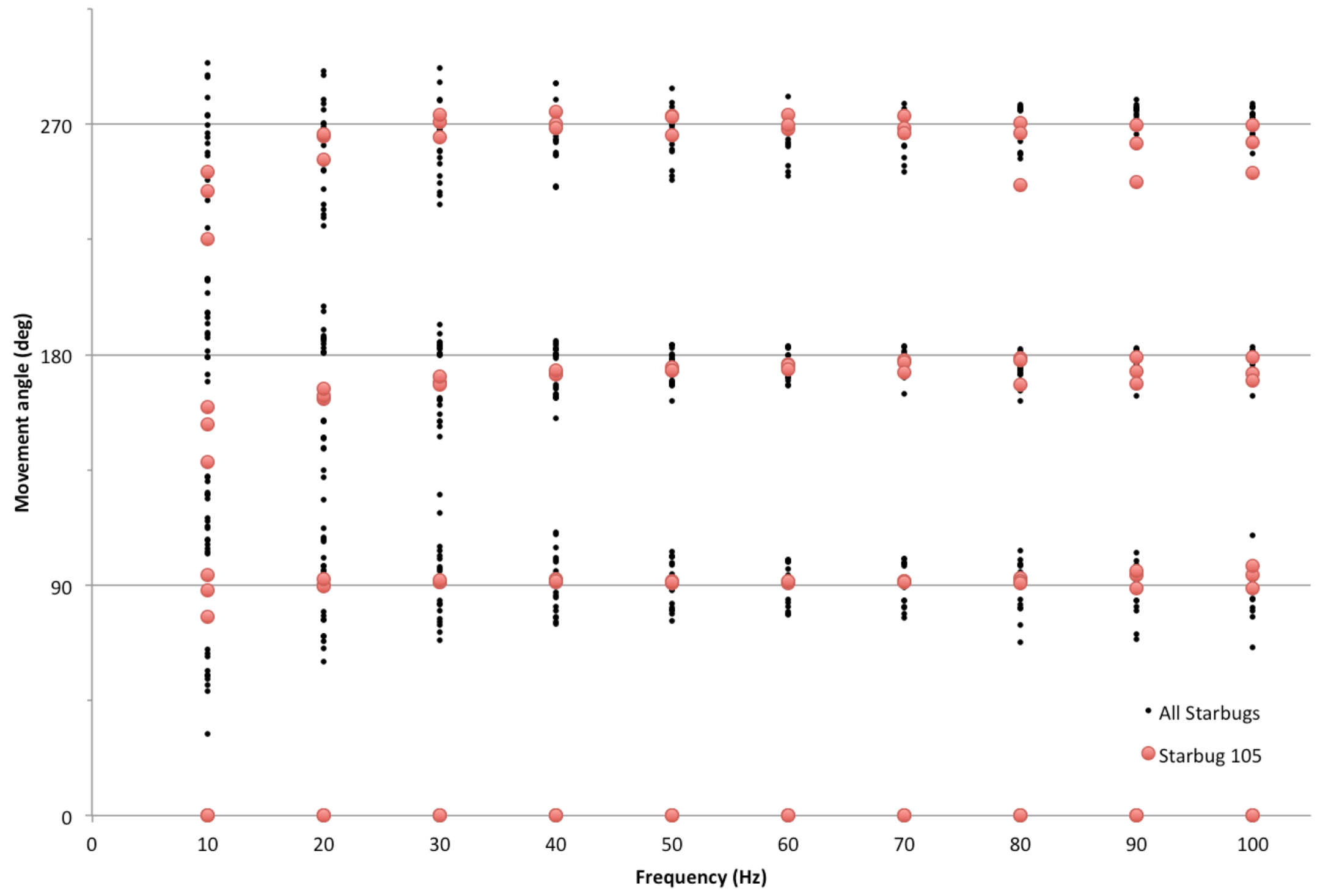}
\end{center}
\caption[personalityFreqAnglePlot] 
{ \label{fig:personalityFreqAnglePlot} 
Starbug personalities: the plot shows a test of 30~Starbugs (Starbug ID=105 is highlighted in red to guide the eye). For each of the 4 movement directions ($\pm x, \pm y$) the angular variation from ideal is shown, as a function of driving frequency. Starbug motions show larger deviations from ideal at low and high drive frequencies, so the Starbugs are predominantly driven at frequencies $\geq 50$~Hz.
}
\end{figure}

Figure~\ref{fig:sbPersonalities} shows the free motion of 29 Starbugs under closed-loop control,
initially pointing in random orientations and commanded to move 20mm north (left-hand plot) or south (right-hand plot) of their current positions, using translational motion only (i.e.\ no rotation) and no calibration. 
Each Starbug exhibits a characteristic behaviour, which is governed by its position, the proximity and location of its neighbours, and the presence of dust on the plate.
A Starbug under ideal conditions should show a direct saw-tooth path from its original position to the commanded destination (shown by a green dot in the figure). Some Starbugs in the test showed something close to this - e.g.\ ID=20, 40, 58 in the first (left) test, and ID=2, 5, 40 in the second (right) test.
Looking at the personalities more closely (Figure~\ref{fig:personalityFreqAnglePlot}) we can see that they also vary as a function of the frequency and voltage at which the Starbug is driven.

Because the Starbugs' deviations from ideal motion vary unpredictably with position and time, the usual method of calibrating the motion at the start of the observing session is not sufficient. Instead, the Starbugs' motion is continually characterised over the course of the night. 
That is, every motion control/monitor point pair collected during a configuration is potentially used as a calibration point. The difference between the expected and actual motions is measured for every motion segment, and a new set of calibration parameters derived if the motion differs from the previously recorded personality by more than a predetermined amount. 
The 44 main and 26 auxiliary Starbug properties (dubbed their {\em personality}) are stored in a mongoDB database. This allows us to keep track of behavioural changes as the Starbugs age, and also of environmental changes --- e.g.\ about the buildup of dust on the plate which will inform how often cleaning should be carried out.

The reason for continuous calibration of the Starbugs is that we are aiming to decrease the reconfiguration time to as short as possible. If a fraction of the closed-loop iterations can be avoided by using our knowledge of how a Starbug has moved within the current parameters, then configuration time may be minimised.

\subsection{Instrument Control}
The TAIPAN instrument is designed to operate autonomously, with little or no local or remote user interaction. Thus high importance is place on the safety of the instrument and the relatively-fragile Starbugs in particular, as there will be no astronomer on duty to respond to alarms from the system during the course of the night.

A Master Controller starts and shuts down the software system and ensures that all components are operational. It contains the high-level intelligence required to coordinate actions between the various software subsystems and to ensure instrument safety. This DRAMA~2 control task also brokers requests from JEEVES, the science scheduler, and returns data, monitor, and error information via the DRAMA/python interface (see Figure~\ref{fig:systemDiagram}).

The Positioner and Spider subsystems are responsible for closed-loop control and monitoring of the 150~Starbugs. The Spider deploys the Starbugs onto the field plate, and ensures that vacuum pressure is maintained and the various mechanisms are correctly positioned for observation. The Positioner determines the current plate location of each Starbug, and moves it to its next target position based on either a pre-determined trajectory or calculates one in real time. Field reconfiguration is carried out in closed-loop control, with the Metrology module providing position updates on the Starbugs at a rate of 4~Hz, while they are in motion until the Starbugs are within the target tolerance.

\begin{figure} [ht]
\begin{center}
\begin{tabular}{c} 
\includegraphics[width=\linewidth]{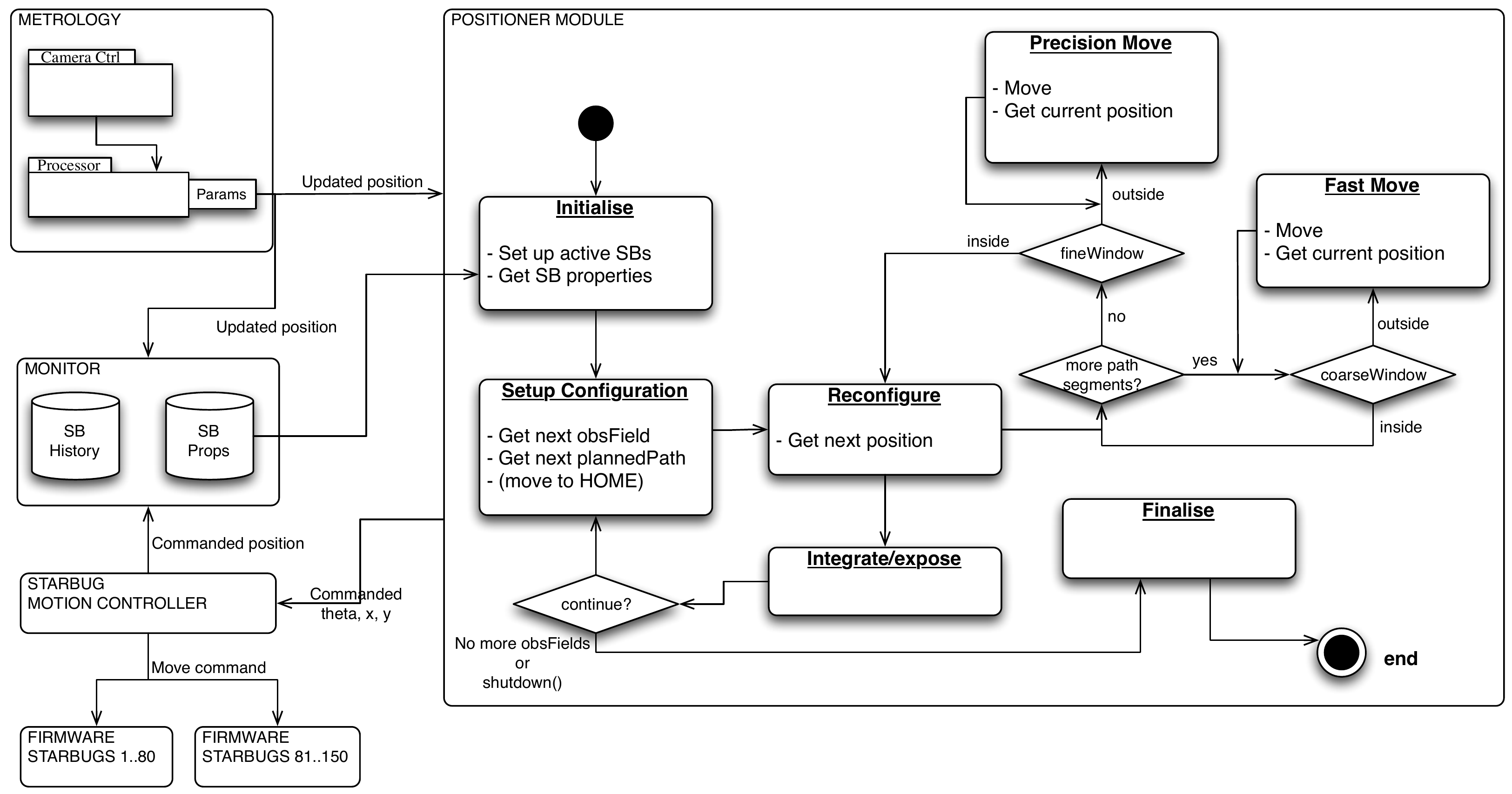}
\end{tabular}
\end{center}
\caption[positionerModuleDesign] 
{ \label{fig:positionerModuleDesign} 
A block diagram representing the operation of the Positioner module. The set of active Starbugs is initialised using the most recent properties stored in the {\em monitor} database. The requested field configuration is obtained from the JEEVES scheduler (via the Master Controller) and the next path segment is determined for each Starbug. The Starbugs are commanded to move to each waypoint at high speed ({\em Fast Move}) until no more path segments remain, at which point the movement mode changes to a slower speed, for maximum precision ({\em Precision Move}). Finally each Starbug is directed towards its target until its metrology position is within the {\em fineWindow} tolerance of the target. This process continues until there are no more fields to be observed or a {\em shutdown} command is received.}
\end{figure} 

\subsection {Path Planning}
For successful field configuration and optimum use of all the Starbugs a collision-free path must be found for each deployed Starbug, which minimises the travel time to target, the interference or tangling between the umbilicals of adjacent Starbugs, and the angle of rotation subtended by a Starbug (to minimise focal ratio degradation in the science fibre). Additionally, the position and path assigned to a given Starbug must not render another's target position unreachable. 

As determining the optimum path for 300~Starbugs can be computationally expensive for some reconfigurations, path planning has been designed to be carried out online, during the field reconfiguration, or off-line (e.g. during the day) facilitating a set of observations to be set up ahead of time.
The off-line process determines 3 paths for each Starbug:
\begin{enumerate}
\item the path between the current and next target,
\item the path between the current target and the canonical {\em home} position (the position at which the ideal Starbug deployment would place the Starbugs on the field plate), and
\item the path between the {\em home} position and the next target.
\end{enumerate}
Because a pre-calculated path to {\em home} always exists, recovering from a fault or omitting one or more fields (e.g. due to bad weather) can be carried out quickly and with a minimum of path recalculation.

Although the Starbugs are independently controlled, the movement mode (i.e. rotation or translation) of each group of 80 Starbugs must be the same at a given time. (This is due to constraints imposed by the instrument's electronics). To manage this the Router module subdivides the reconfiguration into a series of {\em ticks} --- each defined as a movement interval, whose length is defined by the Starbug in the group which will take the longest to reach its next waypoint.  During a {\em tick} the entire group of Starbugs will either rotate on their central axes, translate, or wait until the end of the {\em tick}.

The path planning process consists of three stages of increasing complexity for determining a valid path for each Starbug\cite{2014SPIE.9152E..0SG}. If a path cannot be found using the earlier, computationally cheaper methods, more complex and expensive methods are attempted, until a valid path is found or the target is deemed unreachable.

The three stages currently used are as follows:
\begin{enumerate}
\item {\bf Simple vector}: The Simple Vector is a priority-based routing algorithm. It detects possible path crossings and collisions between pairs of Starbugs paths in the current {\em tick}, and prioritises those Starbugs with no interactions, calculating their paths first. Next each crossing pair is attempted, with one of the Starbugs in the pair being calculated first and then the other. Finally the path for any remaining bugs is attempted. Any Starbugs for which a valid path could not be found proceed to the next stage.

\item {\bf Traffic lights}: This algorithm introduces the option of directing a Starbug to pause for the duration of one or more {\em ticks} to allow nearby Starbugs to proceed, thus avoiding a collision or entanglement.

\item {\bf Traffic lights \& Cooperative A$^*$}: The simple A$^*$ algorithm\cite{1968hnr} determines the least-cost path between two points using a combination of a geometric (past) cost and a heuristic (predicted future) cost. It is grid-based (Starbugs require continuous positioning on the field plate) and cannot handle moving obstacles (e.g.\ neighbouring Starbugs). The Cooperative A$^*$ algorithm, however, modifies this by adding a time dimension to the grid. The grid is now 4-dimensional and consists of the two cartesian axes, a {\em tick} axis, and a reserved path axis. Using Cooperative A$^*$ in conjunction with the Traffic Lights algorithm a Starbug can reserve a path through the grid for a given {\em tick} and thus avoid the paths of other Starbugs. 
\end{enumerate}

Simulation has shown this path planning strategy to be successful in a large number of use-cases, but some test configurations remain unsolved by this 3-stage approach. Work on further refining our approach continues, with the aim of avoiding deadlocks and improving the overall Starbug reconfiguration efficiency.

\section{CONCLUSION}
The AAO has developed the Starbug positioner technology for a new generation of rapid, parallel, fibre positioning instruments. The construction of TAIPAN, a 150-fibre Starbug positioner and dedicated spectrograph is nearing completion and is scheduled to ship to the UK-Schmidt telescope at Siding Spring Observatory in the final quarter of 2016, with a plan to upgrade the instrument to 300 Starbugs in 2017. Further, TAIPAN is a prototype instrument for MANIFEST on the GMT.

Central to the operation of the TAIPAN instrument is the software, which controls the hardware and interfaces to the science user - in our case the autonomous JEEVES scripts. Further, it orchestrates the movement of the independently controlled Starbugs, ensuring fast reconfigurations which are collision and tangle free, maximising the instrumental efficiency for the benefit of the Taipan and FunnelWeb surveys.

\bibliography{main} 
\bibliographystyle{spiebib} 

\end{document}